\documentstyle[epsfig]{aprim}

\newif\ifAMStwofonts

\ifoldfss
  \ifCUPmtlplainloaded \else
    \NewTextAlphabet{textbfit} {cmbxti10} {}
    \NewTextAlphabet{textbfss} {cmssbx10} {}
    \NewMathAlphabet{mathbfit} {cmbxti10} {} 
    \NewMathAlphabet{mathbfss} {cmssbx10} {} 
  \fi
  \ifAMStwofonts
    \ifCUPmtlplainloaded \else
      \NewSymbolFont{upmath} {eurm10}
      \NewSymbolFont{AMSa} {msam10}
      \NewMathSymbol{\upi}     {0}{upmath}{19}
      \NewMathSymbol{\umu}     {0}{upmath}{16}
      \NewMathSymbol{\upartial}{0}{upmath}{40}
      \NewMathSymbol{\leqslant}{3}{AMSa}{36}
      \NewMathSymbol{\geqslant}{3}{AMSa}{3E}

    \fi
  \fi
\fi 

\ifnfssone
  \newmathalphabet{\mathit}
  \addtoversion{normal}{\mathit}{cmr}{m}{it}
  \addtoversion{bold}{\mathit}{cmr}{bx}{it}
  \newmathalphabet{\mathbfit} 
  \addtoversion{normal}{\mathbfit}{cmr}{bx}{it}
  \addtoversion{bold}{\mathbfit}{cmr}{bx}{it}
  \newmathalphabet{\mathbfss} 
  \addtoversion{normal}{\mathbfss}{cmss}{bx}{n}
  \addtoversion{bold}{\mathbfss}{cmss}{bx}{n}
  \ifAMStwofonts
    \ifCUPmtlplainloaded \else
      \UseAMStwoboldmath
      \makeatletter
      \new@mathgroup\upmath@group
      \define@mathgroup\mv@normal\upmath@group{eur}{m}{n}
      \define@mathgroup\mv@bold\upmath@group{eur}{b}{n}
      \edef\UPM{\hexnumber\upmath@group}
      \new@mathgroup\amsa@group
      \define@mathgroup\mv@normal\amsa@group{msa}{m}{n}
      \define@mathgroup\mv@bold\amsa@group{msa}{m}{n}
      \edef\AMSa{\hexnumber\amsa@group}
      \makeatother
      \mathchardef\upi="0\UPM19
      \mathchardef\umu="0\UPM16
      \mathchardef\upartial="0\UPM40
      \mathchardef\leqslant="3\AMSa36
      \mathchardef\geqslant="3\AMSa3E
    \fi
  \fi
\fi 

\ifnfsstwo
  \DeclareMathAlphabet{\mathbfit}{OT1}{cmr}{bx}{it}
  \SetMathAlphabet\mathbfit{bold}{OT1}{cmr}{bx}{it}
  \DeclareMathAlphabet{\mathbfss}{OT1}{cmss}{bx}{n}
  \SetMathAlphabet\mathbfss{bold}{OT1}{cmss}{bx}{n}
  \ifAMStwofonts
    \ifCUPmtlplainloaded \else
      \DeclareSymbolFont{UPM}{U}{eur}{m}{n}
      \SetSymbolFont{UPM}{bold}{U}{eur}{b}{n}
      \DeclareSymbolFont{AMSa}{U}{msa}{m}{n}
      \DeclareMathSymbol{\upi}{0}{UPM}{"19}
      \DeclareMathSymbol{\umu}{0}{UPM}{"16}
      \DeclareMathSymbol{\upartial}{0}{UPM}{"40}
      \DeclareMathSymbol{\leqslant}{3}{AMSa}{"36}
      \DeclareMathSymbol{\geqslant}{3}{AMSa}{"3E}
    \fi
  \fi
\fi 

\ifCUPmtlplainloaded \else
  \ifAMStwofonts \else 
    \def\upi{\pi}
    \def\umu{\mu}
    \def\upartial{\partial}
  \fi
\fi

\title[]{The colliding-wind binary WR140: \\the particle acceleration
laboratory}

\author[Dougherty \& Pittard]
       {S.M.~Dougherty$^1$ and J.M.~Pittard$^2$\\
        $^1$National Research Council, Herzberg Institute for
        Astrophysics, Dominion Radio Astrophysical Observatory,
        Canada\\ 
        $^2$School of Physics \& Astronomy, University of
        Leeds, Leeds, UK}
\date{}

\pagerange{\pageref{firstpage}--\pageref{lastpage}}
\pubyear{2005}

\begin{document}

\maketitle

\label{firstpage}

\begin{abstract}
WR+O star binary systems exhibit synchrotron emission arising from
relativistic electrons accelerated where the wind of the WR star and
that of its massive binary companion collide - the wind-collision
region (WCR). These ``colliding-wind'' binaries (CWB), provide an
excellent laboratory for the study of particle acceleration, with the
same physical processes as observed in SNRs, but at much higher mass,
photon and magnetic energy densities. WR140 is the best studied 
CWB, and high resolution radio observations permit a
determination of several system parameters, particularly orbit
inclination and distance, that are essential constraints for newly
developed models of CWBs. We show a model fit to the radio data at
orbital phase 0.9, and show how these models may be used to predict
the high energy emission from WR140.
\end{abstract}


\section{Introduction}
The 7.9-year period WR+O system WR 140 is the archetype CWB. It
exhibits dramatic variations in its emission from near-IR to radio
wavelengths (Williams et al. 1990; White \& Becker 1995), and at X-ray
energies (Pollock et al. 2005), that are modulated by the highly
eccentric orbit ($e\approx0.88$), with the stellar separation varying
between 2 and 30~AU.  The variations in the synchrotron emission could
arise due to a number of mechanisms. The most widely discussed
possibility is changing free-free opacity along the line-of-sight
through the stellar wind envelopes to the WCR as the orbit progresses.
However, large changes in the stellar separation ($D$) in an orbit
like WR140 alters the intrinsic synchrotron luminosity of the WCR
($\propto D^{-1/2}$) and the free-free and synchrotron absorption
within the WCR. The Razin effect and Coulomb cooling all increase with
decreasing $D$. Inverse-Compton (IC) cooling of the shocked gas is
also important, particularly at higher frequencies, varying strongly
with separation.  All of these processes are included in recently
developed models of CWBs that are based on 2-D hydrodynamical models
to describe both the stellar winds and the WCR, giving a more accurate
representation of the spatial distribution of the thermal and
non-thermal emission (Dougherty et al. 2003; Pittard et al. 2005).

\section{Observations}
\begin{figure}
 \begin{center}
   \epsfig{file=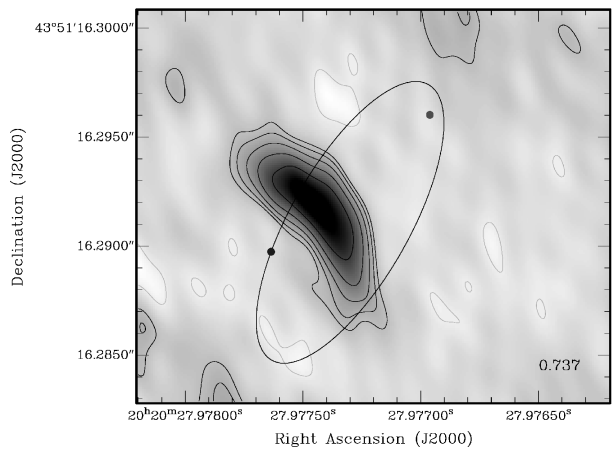,width=8.0cm}
 \end{center}
 \caption{An 8.4~GHz VLBA observation of WR\thinspace140 at orbital
 phase 0.737, with the deduced orbit superimposed. The WR star is to
 the NW.\label{fig1}}
\end{figure}

\begin{figure}
 \begin{center}
   \epsfig{file=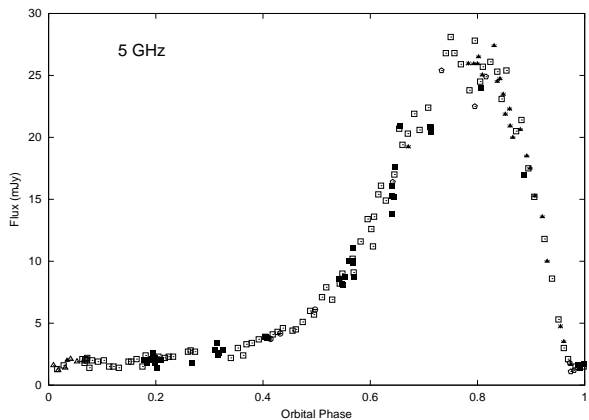,width=8.0cm}
 \end{center}
 \caption{ 5~GHz VLA observations of WR\thinspace140 as a function of
 orbital phase from orbit cycles between 1978-1985 (pentagons),
 1985-1993 (squares), 1993-2001 (triangles), and the current cycle
 2000-2007 (circles). Open symbols are from the VLA, and solid symbols
 from the WSRT.\label{fig2}}
\end{figure}

In addition to new models, recent observations have been obtained with
the VLA and the VLBA that give new constraints to models of WR140.  A
24-epoch campaign of VLBA observations of WR140 was carried out
between orbital phase 0.7 and 0.9. An arc of emission is observed,
resembling the bow-shaped morphology expected for the WCR
(Fig.~\ref{fig1}). This arc rotates from ``pointing'' NW to W as the
orbit progresses which, in conjunction with the observed separation
and position angle of the two stellar components at orbital phase 0.3
(Monnier et al. 2004), leads to solutions for the orbit inclination of
$122\pm5^\circ$, the longitude for the ascending node of
$353\pm3^\circ$, and the orbit semi-major axis of
$9.0\pm0.5$~mas. From the $a\sin i$ derived by Marchenko et
al. (2003), we can derive a distance of $1.85\pm0.16$~kpc to
WR\thinspace140. This represents the first distance derived for CWB
systems {\em independent} of stellar parameters, and together with the
optical luminosity of the system implies the O star is a supergiant.
In addition, total flux measurements from the VLA (Fig.~\ref{fig2})
show that the radio variations from WR\thinspace140 are very closely
the same from one orbit to the next, pointing strongly toward
emission, absorption and cooling mechanisms that are controlled
largely by the orbital motion (Dougherty et al. 2005).

\section{Modelling the radio emission}
\begin{figure}
 \begin{center}
   \epsfig{file=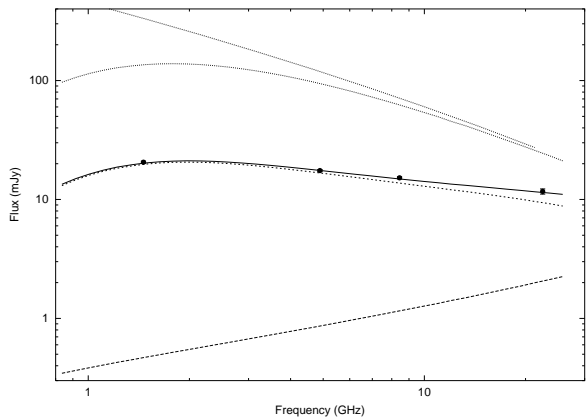,width=8.0cm}
 \end{center}
\caption{The spectrum of WR140 at orbital phase 0.9, fitted with a
model with a wind-momentum ratio of 0.11. The observations are the
solid circles. Various emission components from the model are shown -
free-free (long-dashed), synchrotron flux (short-dashed), intrinsic
synchrotron flux (dotted), and total flux (solid). The top curve shows
the intrinsic synchrotron spectrum without Razin effect.
\label{fig3}}
\end{figure}

Using these new system parameters, we have applied newly developed
radiative transfer models of CWBs to the spectrum of WR\thinspace140
in order to investigate the emission and absorption processes that
govern the radio variations. At orbital phase 0.9, an excellent fit to
the spectrum is possible (Fig.~\ref{fig3}). The free-free flux is
negligible compared to the synchrotron flux, which suffers a large
amount of free-free absorption by the unshocked O-star wind, as
anticipated at this orbital phase since the bulk of the WCR is
'hidden' behind the photospheric radius of the O-star wind. The low
frequency turnover in this model is due to the Razin effect. Similar
fits can be determined for the spectrum at phase 0.8. However, we have
difficulty at earlier orbital phases ($\sim0.4$) if the low frequency
turnover is the result of the Razin effect, largely due to the low
value of magnetic field strength that is required, and the
commensurate extremely high acceleration efficiency that is
implied. We are continuing to investigate this issue.

\section{Future directions}
Our models of the radio emission provide both the spatial distribution
and population of non-thermal electrons. From these, a robust estimate
of the high energy emission at X-ray and $\gamma$-ray energies is
possible. This is of interest due to the potential association between
WR140 and the EGRET source 3EG J2022+4317 (Romero, Benaglia \& Torres,
1999), and is particularly relevant with several high energy
satellites currently in orbit (Chandra, XMM-Newton, INTEGRAL) and
further satellites and observatories which will be operational in the
near future (GLAST and VERITAS). Previous work in this area has used
the observed synchrotron luminosity and the equipartition magnetic
field to estimate the IC luminosity (e.g. Benaglia \& Romero
2003). However, a key consequence of our radio modelling is that we
derive the intrinsic synchrotron luminosity, which is almost an order
of magnitude different to the observed luminosity at phase 0.9 (Fig~\ref{fig3}),
and we determine the spatial distribution of the magnetic field.  In a
forthcoming paper, we explore the IC emission, relativistic
bremsstrahlung, pion-decay processes and absorption due to
pair-production in WR140, and the other well-studied CWBs WR146 and
WR147 (Pittard \& Dougherty, 2005).

\label{lastpage}

\clearpage


\begin{thebibliography}{99}
\bibitem[\protect\citeauthoryear{Benaglia \& Romero}{2003}]{Benaglia:2003}
Benaglia P., Romero G.~E. 2003, A\&A, 399, 1121
\bibitem[\protect\citeauthoryear{Dougherty et al.}{2005}]{Dougherty:2005}
Dougherty S.~M., Beasley A.~J., Claussen M.~J., Zauderer B.~A., Bolingbroke N.~J. 2005, ApJ, 623, 447
\bibitem[\protect\citeauthoryear{Marchenko et~al.}{2003}]{Marchenko:2003}
{Marchenko} S.~V. et al. 2003, ApJ, 596, 1295
\bibitem[\protect\citeauthoryear{Monnier et~al.}{2004}]{Monnier:2004}
Monnier J.~D. et al. 2004, ApJL, 602, L57
\bibitem[\protect\citeauthoryear{Pittard \& Dougherty}{2005}]{Pittard:2005a}
Pittard J.~M., Dougherty S.~M. 2005, A\&A, submitted
\bibitem[\protect\citeauthoryear{Pittard et al.}{2005}]{Pittard:2005b}
Pittard J.~M., Dougherty S.~M., Coker R.~F., O'Connor E., Bolingbroke N.~J.
2005, A\&A, submitted
\bibitem[\protect\citeauthoryear{Pollock et al.}{2005}]{Pollock:2005}
Pollock A.~M.~T., Corcoran M.~F., Stevens I.~R, Williams P.~M. 2005, ApJ, 629, 482
\bibitem[\protect\citeauthoryear{Romero, Benaglia \& Torres}{1999}]{Romero:1999}
Romero G.~E., Benaglia P., Torres D.~F. 1999, A\&A, 348, 868
\bibitem[\protect\citeauthoryear{White \& Becker}{1995}]{White:1995}
White R.~L., Becker R.~H. 1995, ApJ, 451, 352
\bibitem[\protect\citeauthoryear{Williams et al.}{1990}]{Williams:1990}
Williams P.~M., van der Hucht K.~A., Pollock A.~M.~T., Florkowski D.~R., 
van der Woerd H., Wamsteker W.~M. 1990, MNRAS, 243, 662

\end{thebibliography}
\end{document}